\documentclass{article}
\usepackage{spconf,amsmath,graphicx,hyperref,booktabs,amssymb}
\usepackage[table]{xcolor}

\usepackage[dvipsnames]{xcolor}
\definecolor{navy}{RGB}{30, 50, 150}

\title{SelectSep: Target-Selective Video Queries for Audio Source Separation}
\name{Suyeon Lee, Doyeop Kwak, Joon Son Chung}
\address{Korea Advanced Institute of Science and Technology, South Korea}
\begin{document}
%
\maketitle

\begin{abstract}
Video-queried sound separation (VQSS) aims to isolate a target sound from an audio mixture using its corresponding video as a query. However, scene-level video queries can become ambiguous when multiple sounding objects are visible, particularly when the target and interferer are semantically similar. We propose SelectSep, a target-selective VQSS framework that jointly uses text and a single click coordinate to specify the target source and instance without requiring segmentation at inference time. We adapt the visual encoder through target-feature distillation with text and click conditioning, and then fine-tune the separator using the adapted visual features. We also introduce an on-screen interference setting, where both the target and interferer are visible, to evaluate target selectivity. Experiments on standard benchmarks and the on-screen interference setting demonstrate consistent improvements in separation quality and target selectivity, particularly for mixtures containing semantically similar sources.
\end{abstract}


\section{Introduction}
\label{sec:intro}

Universal sound source separation aims to isolate a desired source from an arbitrary mixture using a user-provided query. Thus, precise target specification is essential to extract the desired sound. Various query modalities have been explored, including text~\cite{liu2022separate,audiosep,yuan2025flowsep}, image~\cite{dong2022clipsep,chen2023iquery,cheng2025omnisep}, and video~\cite{cheng2026alignsep,huang2025davisflow,mmaudiosep}. 
Video-queried sound separation (VQSS) is particularly attractive, as video provides not only semantic information but also temporal audio-visual correspondence.
However, conventional VQSS systems may not fully exploit these cues when multiple sounding objects appear in the same scene. When both the target and interferer are visible, scene-level video features may contain information from both sources, leading to ambiguous semantic and synchronization cues.

This ambiguity has motivated approaches that make visual queries more target-specific. In VQSS, VAST~\cite{tan2023vast} incorporates a text query to semantically specify the target in the video, while SAM-Audio~\cite{samaudio} additionally uses object-level regions obtained from segmentation for more explicit target specification. Related approaches have also emerged in video-to-audio generation: SelVA~\cite{selva} fine-tunes the visual encoder with text conditioning for semantic selectivity, while SAGANet~\cite{viertola2026saganet} uses spatial segmentation masks to indicate the target region.  These approaches highlight two complementary forms of selectivity. Semantic conditioning, such as text, specifies \emph{what} source to attend to, but may require increasingly detailed descriptions to disambiguate similar instances in cluttered scenes. Spatial conditioning, such as a segmentation mask, directly specifies \emph{which} instance is intended, but dense masks require an additional segmentation model at inference time. This motivates a visual query that combines semantic selectivity with a simple and direct way to specify the target instance.

\begin{figure*}[t]
    \centering
    \includegraphics[width=0.99\linewidth]{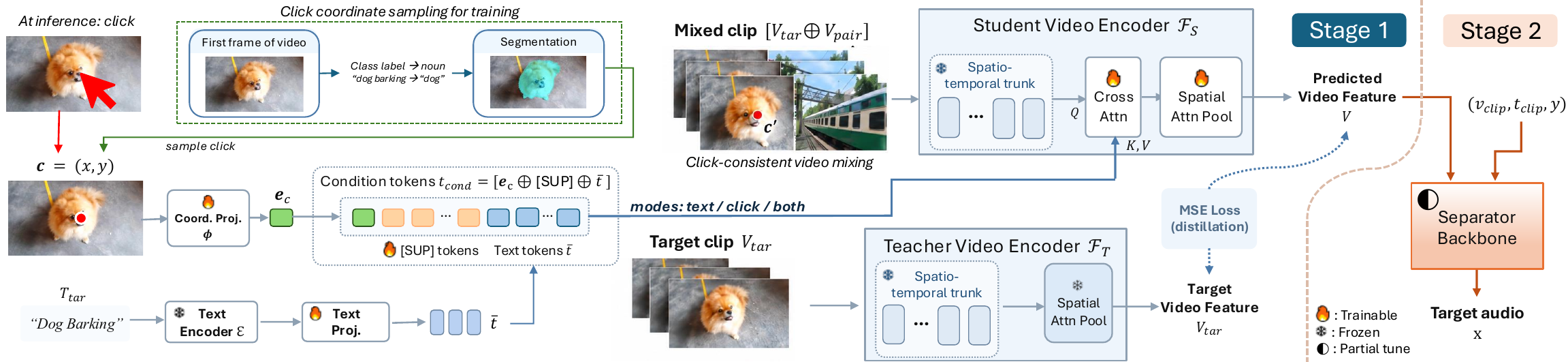}
    \vspace{-2mm}
    \caption{
    Training pipeline of SelectSep. In stage 1, we fine-tune the visual encoder with text and click conditions for target selectivity at the video feature level. In stage 2, we freeze the student video encoder and fine-tune the separator backbone.
    }
    \label{fig:vis_enc_finetune}
\end{figure*}

To this end, we propose \textbf{SelectSep}, which introduces target selectivity into the visual query through joint text and coordinate conditioning. Building on the selective visual encoder fine-tuning strategy in~\cite{selva}, we extend selectivity with lightweight spatial conditioning using a single click. This design allows users to specify the target object directly at inference time without requiring a segmentation model. We further introduce an \emph{on-screen interference} setting to evaluate target selectivity in VQSS, where both the target and interferer are visible in the video. We adapt VGG-MonoAudio, originally proposed for selective video-to-audio generation~\cite{selva}, into a source separation benchmark by mixing the audio tracks corresponding to two concatenated videos. Its intra-class subset, where the target and interferer share the same semantic category, further evaluates instance-level selectivity beyond coarse semantic cues. 
Experimental results across standard benchmarks and the proposed on-screen interference setting validate the effectiveness of the proposed target-selective conditioning.

\section{Method}
\label{sec:method}

\subsection{Stage 1: Click-conditioned visual encoder adaptation}
\label{ssec:click}

We introduce instance-level selectivity into a text-selective visual encoder by conditioning it on a single click coordinate $\mathbf{c}=(x,y)\in[0,1]^2$ in addition to text. Our encoder builds on the text-conditioned Synchformer adaptation of SelVA~\cite{selva}, where lightweight conditioning modules are trained by teacher--student distillation. We retain this framework and modify the conditioning mechanism to jointly support semantic and spatial target specification (Fig.~\ref{fig:vis_enc_finetune}).

\noindent\textbf{Click-conditioned visual encoding.}
Following the text-conditioning design introduced in \cite{selva}, projected text features $\bar{\mathbf{t}}=\mathrm{Proj}(\mathcal{E}(T))$ and learnable $\texttt{[SUP]}$ tokens form the key/value sequence of the text-guided cross-attention module. We map the click coordinate on the target object to the same embedding space using an MLP, $\mathbf{e}_{\mathbf{c}}=\phi(\mathbf{c})$, and prepend it to this sequence: 
$ \mathbf{t}_{\mathrm{cond}}
=
[\,\mathbf{e}_{\mathbf{c}}
\oplus \texttt{[SUP]}
\oplus \bar{\mathbf{t}}\,].$
This allows the visual encoder to jointly use semantic information from text and spatial information from the click. We support three conditioning modes: \textsc{text}, \textsc{click}, and \textsc{both}.

\noindent\textbf{Condition dropout.}
To make either cue usable independently at inference, we apply mutually exclusive condition dropout during training: the text cue is removed with probability $p_{\mathrm{txt}}=0.3$, the click cue with $p_{\mathrm{clk}}=0.15$, and otherwise both are retained. The two cues are never dropped simultaneously, and the $\texttt{[SUP]}$ tokens are always preserved.

\noindent\textbf{Coordinate-consistent video mixing.}
During training, we apply target--distractor video mixing to $p_{\mathrm{mix}}=0.75$ of the training samples, with the target placed on either side of the concatenated video with equal probability. The click is transformed together with the target crop so that it remains aligned with the same object; for a target occupying a width fraction $\lambda$, its horizontal coordinate is rescaled and shifted accordingly, while the vertical coordinate is unchanged.

\noindent\textbf{Training objective.}
A frozen Synchformer teacher $\mathcal{F}_T$ encodes the clean target clip, while the student $\mathcal{F}_S$ receives the mixed video and the available text/click condition. We optimize the student to reproduce the teacher representation:
\begin{equation}
\mathcal{L}_{\mathrm{dist}}
=
\left\|
\mathcal{F}_S\!\left(
[V_{\mathrm{tar}}\oplus V_{\mathrm{pair}}],
\mathbf{t}_{\mathrm{cond}}
\right)
-
\mathcal{F}_T(V_{\mathrm{tar}})
\right\|_2^2 ,
\label{eq:distill_click}
\end{equation}
while spatio-temporal trunk remains frozen; only the cross-attention module, text and coordinate projections, $\texttt{[SUP]}$ tokens, and spatial attention pooling layer are updated.

\subsection{Stage 2: Separator fine-tuning with adapted encoder}
\label{ssec:stage2}

\noindent\textbf{Generator fine-tuning.}
We freeze the Stage 1 encoder $\mathcal{F}_S$ and replace the Synchformer encoder in the MMAudioSep~\cite{mmaudiosep} architecture with it. Starting from pretrained MMAudio-S-16k weights, we fine-tune the multimodal DiT blocks. The synchronization-feature projector is also fine-tuned to accommodate the adapted encoder's features, and we use a single video for visual input.
We retain MMAudioSep's CLIP~\cite{radford2021clip} visual and text conditioning and train with its conditional flow matching objective. The model operates in the VAE latent space of mel-spectrograms, with BigVGAN synthesizing the waveform.

\begin{table*}[t]
\centering
\caption{
Source separation results. T, V, and I denote text, video, and image. $\dagger$ denotes the target-selective video query conditioned on click and text inputs.
Best and second-best scores are bolded and underlined.
}
\vspace{1mm}
\label{tab:available_models_comp}
\renewcommand{\arraystretch}{1.12}
\resizebox{1\textwidth}{!}{%
\begin{tabular}{lccc|cccccc|cccccc}
\toprule
\midrule
& \multicolumn{3}{c|}{Query}
& \multicolumn{6}{c|}{\textbf{VGGSound-Clean}}
& \multicolumn{6}{c}{\textbf{MUSIC-Clean}} \\
\cmidrule(lr){2-4}
\cmidrule(lr){5-10}
\cmidrule(lr){11-16}

Method
& T & V & I
& FAD$\downarrow$
& IS$\uparrow$
& CLAP$\uparrow$
& CLAP-A$\uparrow$
& IB-Score$\uparrow$
& DeSync$\downarrow$
& FAD$\downarrow$
& IS$\uparrow$
& CLAP$\uparrow$
& CLAP-A$\uparrow$
& IB-Score$\uparrow$
& DeSync$\downarrow$ \\
\midrule

\textcolor{gray}{\textit{GT (reconstructed)}} &  & & 
& \textcolor{gray}{2.59}
& \textcolor{gray}{10.00}
& \textcolor{gray}{25.49}
& \textcolor{gray}{88.98}
& \textcolor{gray}{25.89}
& \textcolor{gray}{0.718}
& \textcolor{gray}{4.92}
& \textcolor{gray}{2.78}
& \textcolor{gray}{32.71}
& \textcolor{gray}{92.20}
& \textcolor{gray}{38.74}
& \textcolor{gray}{0.163} \\

FlowSep~\cite{yuan2025flowsep}
& \checkmark & - & -
& \textbf{1.71}
& \textbf{9.86}
& \underline{24.42}
& \underline{81.48}
& 24.86
& 0.844
& 5.42
& \underline{2.73}
& 29.08
& 79.00
& 29.94
& 0.461 \\

OmniSep~\cite{cheng2025omnisep}
& \checkmark & - & \checkmark
& \underline{1.74}
& 9.30
& 22.67
& 81.07
& \textbf{26.31}
& 0.835
& \underline{4.73} 
& 2.45
& \underline{32.29}
& 84.17
& 35.59
& 0.303 \\

DAVIS-Flow~\cite{huang2025davisflow}
& - & \checkmark & -
& 4.59
& 8.40
& 17.06
& 62.62
& 21.68
& 0.950
& 5.72
& 2.64
& 31.83
& 48.61
& 35.31
& 0.374 \\

MMAudioSep~\cite{mmaudiosep}
& \checkmark & \checkmark & -
& 3.54
& 8.59
& 21.71
& 76.88
& 22.51
& \underline{0.769}
& 5.50 
& 2.70 
& 32.10 
& \underline{86.54} 
& \underline{37.68} 
& \underline{0.232} \\

\rowcolor{cyan!10}
\textbf{SelectSep (Ours)} 
& \checkmark & \checkmark$\dagger$ & -
& 2.83
& \underline{9.85}
& \textbf{25.28}
& \textbf{83.91}
& \underline{26.09}
& \textbf{0.756}
& \textbf{3.82}
& \textbf{2.87}
& \textbf{33.59}
& \textbf{90.23}
& \textbf{39.67}
& \textbf{0.212} \\

\midrule
\midrule

& \multicolumn{3}{c|}{Query}
& \multicolumn{6}{c|}{\textbf{VGG-MonoAudio Inter-class}}
& \multicolumn{6}{c}{\textbf{VGG-MonoAudio Intra-class}} \\
\cmidrule(lr){2-4}
\cmidrule(lr){5-10}
\cmidrule(lr){11-16}

Method
& T & V & I
& FAD$\downarrow$
& IS$\uparrow$
& CLAP$\uparrow$
& CLAP-A$\uparrow$
& IB-Score$\uparrow$
& DeSync$\downarrow$
& FAD$\downarrow$
& IS$\uparrow$
& CLAP$\uparrow$
& CLAP-A$\uparrow$
& IB-Score$\uparrow$
& DeSync$\downarrow$ \\
\midrule

\textcolor{gray}{\textit{GT (reconstructed)}} &  & & 
& \textcolor{gray}{7.70}
& \textcolor{gray}{10.17}
& \textcolor{gray}{26.50}
& \textcolor{gray}{89.52}
& \textcolor{gray}{32.92}
& \textcolor{gray}{0.429}
& \textcolor{gray}{3.07}
& \textcolor{gray}{8.00}
& \textcolor{gray}{26.23}
& \textcolor{gray}{90.22}
& \textcolor{gray}{33.89}
& \textcolor{gray}{0.382} \\

FlowSep~\cite{yuan2025flowsep}
& \checkmark & - & -
& 6.93
& \textbf{9.52}
& \textbf{24.00}
& \textbf{79.54}
& \textbf{29.91}
& \textbf{0.532}
& 4.32
& \underline{7.75}
& \underline{22.61}
& 76.39
& 29.22
& \underline{0.576} \\

OmniSep~\cite{cheng2025omnisep}
& \checkmark & - & \checkmark
& \textbf{5.39}
& 8.74
& 20.93
& 76.84
& \underline{28.36}
& 0.582
& \textbf{2.77}
& 7.63
& 21.82
& \underline{78.59}
& \textbf{30.49}
& 0.586 \\

DAVIS-Flow~\cite{huang2025davisflow}
& - & \checkmark & -
& 14.82
& 7.90
& 14.33
& 63.53
& 22.56
& 0.697
& 6.02
& 7.48
& 17.69
& 68.70
& 26.65
& 0.702 \\

MMAudioSep~\cite{mmaudiosep} 
& \checkmark & \checkmark & -
& 15.92
& 7.10
& 16.49
& 63.62
& 20.56
& 0.677
& 5.57
& 6.34
& 18.70
& 70.61
& 24.99
& 0.648 \\

\rowcolor{cyan!10}
\textbf{SelectSep (Ours)} 
& \checkmark & \checkmark$\dagger$ & -
& 7.98
& \underline{9.42}
& \underline{23.76}
& \underline{77.59}
& 28.33
& \underline{0.537}
& \underline{3.61}
& \textbf{7.95}
& \textbf{24.34}
& \textbf{79.40}
& \underline{30.20}
& \textbf{0.488} \\

\bottomrule
\end{tabular}%
}
\vspace{-2mm}
\end{table*}

\noindent\textbf{Mixture-centered prior.}
Following the mixture-centered prior of SGMSE~\cite{richter2023sgmse}, we replace MMAudioSep's zero-centered source distribution with
$q(\mathbf{x}_0\mid\mathbf{y})=\mathcal{N}(\mathbf{y},\sigma^2\mathbf{I})$,
where $\mathbf{y}$ is the normalized mixture latent, and otherwise retain the conditional flow matching objective.

\section{Experimental Setup}
\label{sec:exp_setup}

\subsection{Implementation details}

\noindent\textbf{Data.} The visual encoder is pretrained and fine-tuned solely on VGGSound~\cite{chen2020vggsound}, while separator fine-tuning follows the MMAudioSep setup, using a composition of audio datasets, additionally: AudioCaps~\cite{audiocaps}, Clotho~\cite{clotho}, and the AudioSet-SL, BBCSound, and FreeSound subsets of WavCaps~\cite{mei2023wavcaps}. Two audios are mixed with random SNR in $ [-15, 15] $ dB.

\noindent\textbf{Pseudo-labelling click.} 
We generate click supervision for VGGSound using video masks extracted following~\cite{viertola2026saganet}. For the first frame of each video segment, we draw the click coordinate $\mathbf{c}$ from the inner half of the corresponding mask.

\noindent\textbf{Training.}
Both stages use AdamW ($\beta=(0.9,0.95)$, weight decay $10^{-6}$, gradient clipping $1.0$) with a constant learning rate of $10^{-4}$ after 1k-step linear warmup, bfloat16, and post-hoc EMA.
In Stage 1, we train only the visual encoder's text-conditioned cross-attention and spatial pooling modules for 50k steps, keeping the spatio-temporal trunk frozen.
In Stage 2, we fine-tune the separator with Stage-1 visual features for up to 100k steps with an effective batch size of 512.
We set $\sigma=0.5$ and sample $t\sim\mathcal{U}[0.03,1]$ for mixture-centered prior.

\noindent\textbf{Inference.} Inference uses 25 Euler steps with CFG scale $\gamma=4.5$. We use class labels for VGGSound-based datasets and instrument labels for MUSIC-Clean as text queries.

\subsection{Evaluation}
\noindent\textbf{Standard benchmark.}
We use the publicly released AudioSep~\cite{audiosep} evaluation sets, restricted to the samples whose source video is still retrievable from YouTube. This yields 980 mixtures from VGGSound-Clean~\cite{chen2020vggsound} and
4,943 from MUSIC-Clean~\cite{zhao2018music}, each providing a mixture and its target. Both
sets are mixed at $0$ dB SNR.

\noindent\textbf{On-screen interference.} 
In addition to the existing benchmarks, we adapt VGG-MonoAudio, originally introduced for selective video-to-audio generation~\cite{selva}, to evaluate target selectivity under on-screen interference. The benchmark consists of clean, single-source samples collected from VGGSound~\cite{chen2020vggsound}. It contains 560 inter-class pairs, where the two videos belong to different sound categories, and 511 intra-class pairs, where they belong to the same category. For each pair, the two videos are horizontally concatenated so that both the target and interferer are visible in the query video. We refer readers to~\cite{selva} for details on the original benchmark construction. To adapt this for source separation, we mix the audio tracks of each video pair at $0$ dB SNR and trim to 8 s.

\noindent\textbf{Click annotations.}
We manually annotate ground-truth clicks for VGG-MonoAudio to isolate coordinate conditioning from pseudo-mask errors. As VGGSound and MUSIC do not provide ground-truth masks or clicks, we use pseudo-clicks instead. 
The proportions of samples with available clicks are
82.7\% for VGGSound-Clean, 58.9\% for MUSIC-Clean, and 98.2\% and 99.4\% for VGG-MonoAudio inter-class and intra-class. 
All click coordinates are fixed for reproducibility.
Our visual encoder operates in text-only mode for videos where the target object is not in the first frame.

\noindent\textbf{Baselines.}
We retrain our baseline architecture MMAudioSep~\cite{mmaudiosep} at 16 kHz, and also evaluate separation models using their released checkpoints: 
OmniSep~\cite{cheng2025omnisep}, which supports text, image, and audio queries;  FlowSep~\cite{yuan2025flowsep}, with text queries;  and DAVIS-Flow~\cite{huang2025davisflow}, with video queries. 
OmniSep is trained on VGGSound and MUSIC, while FlowSep follows our audio-data composition except for Clotho. DAVIS-Flow operates at 11.025~kHz; we use its MUSIC-trained checkpoint for MUSIC-Clean and AVE-trained checkpoint otherwise. For OmniSep, we use image and text queries.

\noindent\textbf{Metrics.}
We follow the evaluation protocol of MMAudioSep~\cite{mmaudiosep} using the \texttt{av-benchmark} toolkit and report FAD, Inception Score (IS), CLAP/CLAP-A~\cite{wu2023clap}, ImageBind score (IB-Score)~\cite{girdhar2023imagebind}, and DeSync~\cite{iashin2024synchformer}.

\section{Results}

\subsection{Standard benchmark}
We evaluate SelectSep on VGGSound-Clean and MUSIC-Clean against prior methods (upper half of Tab.~\ref{tab:available_models_comp}). We also report GT (reconstructed) metrics obtained by passing ground-truth audio through the same mel-spectrogram, VAE, and vocoder pipeline used by SelectSep and MMAudioSep, providing a reference for reconstruction effects.

On VGGSound-Clean, SelectSep achieves the best performance among the compared methods on CLAP, CLAP-A, and DeSync, and the second-best on IS and IB-Score. Compared with MMAudioSep, which shares the same backbone, SelectSep performs better across all six metrics. These results indicate that the proposed target-selective adaptation remains effective in the standard setting, beyond the on-screen interference scenarios considered in our evaluation. On MUSIC-Clean, SelectSep outperforms all compared methods on every metric
despite being trained without MUSIC data. MMAudioSep is also trained without MUSIC, whereas OmniSep and DAVIS-Flow include MUSIC in their training data. These results suggest that SelectSep generalizes beyond VGGSound to an unseen audio-visual domain. Notably, GT DeSync is substantially lower on MUSIC-Clean than on VGGSound-Clean, indicating stronger audio-visual temporal alignment in the instrument videos. Such synchronization may particularly favor SelectSep, which is trained to focus on target-specific visual information.

\begin{table}[t]
\centering
\vspace{-2mm}
\caption{Ablation study on VGG-MonoAudio. Components are added cumulatively.}
\label{tab:vgg-mono-ablation}
\renewcommand{\arraystretch}{1.15}
\resizebox{\columnwidth}{!}{%
\begin{tabular}{lcccccc}
\toprule
Method
& FAD$\downarrow$
& IS$\uparrow$
& CLAP$\uparrow$
& CLAP-A$\uparrow$
& IB-Score$\uparrow$
& DeSync$\downarrow$ \\

\midrule
\midrule
\multicolumn{7}{c}{\textbf{\textit{Inter-class}}} \\
\midrule

MMAudioSep~\cite{mmaudiosep}
& 15.92
& 7.10
& 16.49
& 63.62
& 20.56
& 0.677 \\

+ Mixture-centered prior 
& 16.35
& 7.24
& 16.76
& 63.46
& 22.96
& 0.647 \\

+ Text
& 10.60
& 9.25
& 22.85
& 74.17
& 27.30
& 0.588 \\

\textbf{+ Click (Ours)}
& \textbf{7.98}
& \textbf{9.42}
& \textbf{23.76}
& \textbf{77.59}
& \textbf{28.33}
& \textbf{0.537} \\

\midrule
\midrule
\multicolumn{7}{c}{\textbf{\textit{Intra-class}}} \\
\midrule

MMAudioSep~\cite{mmaudiosep}
& 5.57
& 6.34
& 18.70
& 70.61
& 24.99
& 0.648 \\

+ Mixture-centered prior 
& 5.14
& 6.74
& 20.11
& 71.02
& 27.68
& 0.573 \\

+ Text 
& 4.12
& 7.59
& 22.85
& 75.72
& 29.03
& 0.558 \\

\textbf{+ Click (Ours)}
& \textbf{3.61}
& \textbf{7.95}
& \textbf{24.34}
& \textbf{79.40}
& \textbf{30.20}
& \textbf{0.488} \\

\bottomrule
\end{tabular}%
}
\vspace{-3mm}
\end{table}


\subsection{On-screen interference: VGG-MonoAudio}
VGG-MonoAudio contains both the target and interferer on-screen, making target identification challenging for models that rely on scene-level video queries. We therefore evaluate how accurately each model identifies and separates the intended target (bottom half of Tab.~\ref{tab:available_models_comp}).

\noindent\textbf{Inter-class.}
In the inter-class subset, the target and interferer belong to different semantic categories. Since the two sources are semantically distinct, class-level text queries can distinguish them without relying on the visual scene. Accordingly, the text-only FlowSep performs strongly on most metrics, while video-queried models generally show lower performance. MMAudioSep also uses a text query together with video, but its lower performance suggests that the additional scene-level visual query may not fully resolve the ambiguity introduced by the on-screen interferer. SelectSep achieves better results than MMAudioSep across all six metrics, showing the potential benefit of target-selective visual conditioning under on-screen interference. 
FlowSep nevertheless remains better across the reported metrics, leaving room for improvement in handling on-screen visual interference.

\noindent\textbf{Intra-class.}
In the intra-class subset, the target and interferer belong to the same semantic category, making class-level text queries less discriminative for distinguishing the two instances. Scene-level video queries likewise provide limited cues for instance-level disambiguation. SelectSep records the highest IS, CLAP, and CLAP-A, as well as the lowest DeSync, among the compared methods. The lower DeSync is consistent with closer temporal alignment between the separated audio and the target video. SelectSep also performs better than the text-only FlowSep across the reported metrics. These results suggest that target-specific visual conditioning may provide useful additional cues when semantic information alone is insufficient to identify the intended source.


\begin{figure}
\centering
\includegraphics[width=\linewidth]{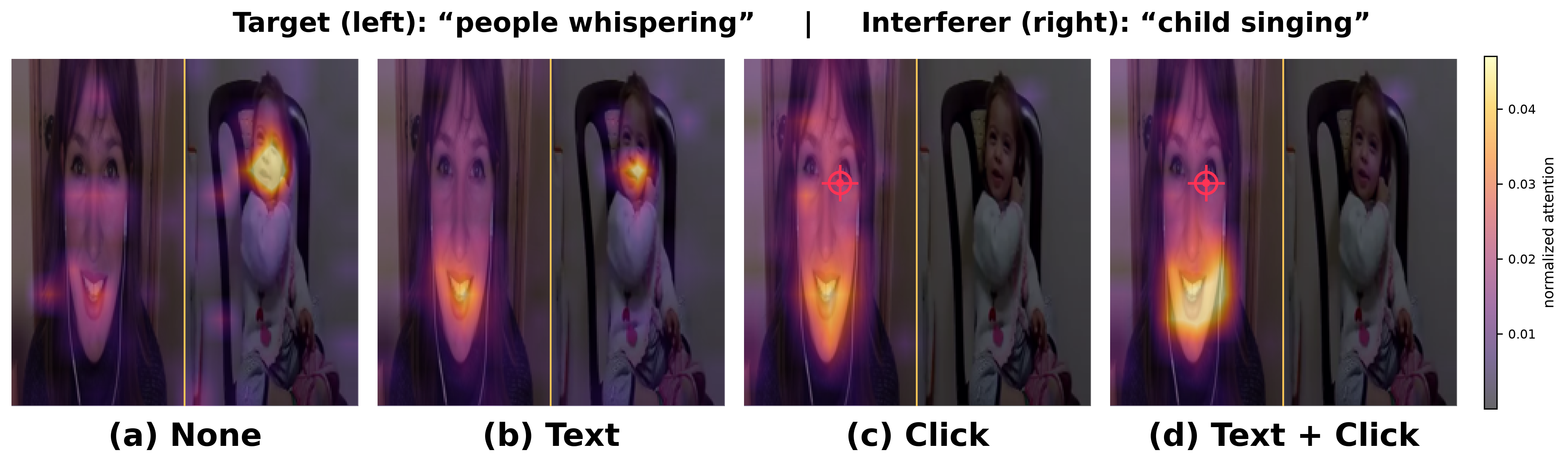}
\vspace{-8mm}
\caption{
Attention maps of the visual encoder under each conditioning mode.
The marker indicates the clicked position.}
\label{fig:attn_visualize}
\vspace{-2mm}
\end{figure}

\subsection{Ablation study}
\noindent\textbf{Effect of model components.}
We present component-wise ablations on VGG-MonoAudio using ground-truth click annotations to isolate the effect of coordinate conditioning from errors in the pseudo masks (Tab.~\ref{tab:vgg-mono-ablation}). Replacing the zero-centered Gaussian prior with the mixture-centered prior improves most metrics, while slightly worsening FAD and CLAP-A on the inter-class subset. Adding text-based target conditioning leads to improvements over the mixture-centered prior, and click conditioning improves all six metrics in both subsets. In particular, the reduction in DeSync with click conditioning is consistent with closer temporal alignment between the separated audio and the target video.

\vspace{1mm}
\noindent\textbf{Qualitative attention analysis.}
To qualitatively examine target selectivity in the fine-tuned visual encoder, we visualize attention under semantically similar on-screen interference. Fig.~\ref{fig:attn_visualize} shows an intra-class sample from VGG-MonoAudio with the attention maps overlaid on the frame. For the vanilla Synchformer~\cite{iashin2024synchformer}, we visualize the self-attention of the last block; for the fine-tuned encoders, we visualize the last-block attention of the \texttt{[eos]} token over the concatenated frame.
The frame contains a woman and a child from the same \textit{human} category, with the woman as the target. The vanilla encoder attends mainly to the child's face, while text conditioning distributes attention over the mouths of both people. Click conditioning shifts the attention toward the target, and combining text and click cues further concentrates it around the woman's lips. These attention patterns are consistent with the quantitative results and provide qualitative evidence for the complementary roles of text and click cues.

\section{Conclusion}
\label{sec:conclusion}

We presented SelectSep, a target-selective VQSS method that uses text and a single click to specify the intended source. By combining semantic and spatial conditioning in the visual encoder, SelectSep enables instance-level target specification without segmentation at inference time. Experiments on standard benchmarks and under on-screen interference show consistent benefits, including for mixtures of semantically similar sources. 
Click conditioning further improves separation and temporal audio-visual alignment beyond text conditioning alone. Overall, these results support the use of complementary semantic and spatial cues for target-selective VQSS.

\vfill\pagebreak

\bibliographystyle{IEEEbib}
\bibliography{strings,refs}

\end{document}